\documentclass[mathpazo]{cicp}

\usepackage{esint}
\usepackage{amsmath,amsthm}
\usepackage{float}
\usepackage{graphicx}
\usepackage{epstopdf}
\usepackage{caption}
\usepackage{subfigure} 
\usepackage{tabularx}
\usepackage{threeparttable}
\usepackage{mathrsfs}

\usepackage{dcolumn}
\usepackage{bm}
\usepackage{color,epsfig,multirow}
\usepackage{array}

\usepackage{hyperref}
\usepackage{cleveref}
\usepackage{epstopdf}
\usepackage{algorithm}
\usepackage{algorithmic}

\usepackage{url}
\allowdisplaybreaks[4]

\begin{document}
\title{Harmonic surface mapping algorithm for electrostatic potentials in an atomistic/continuum hybrid model for electrolyte solutions}

 \author[Fu J et.~al.]{Jing Fu\affil{1,}
       Zecheng Gan\affil{2,}\corrauth}
 \address{\affilnum{1}\ School of Mathematical Sciences, Shanghai Jiao Tong University, Shanghai 200240, China.\\
           \affilnum{2}\ Department of Mathematics, University of Michigan, Ann Arbor, Michigan 48109, U.S.A.  }
\emails{{\tt 179137007@sjtu.edu.cn} (J. Fu), {\tt zecheng@umich.edu} (Z. Gan) }    

\begin{abstract}
Simulating charged many-body systems has been a computational demanding task due to the long-range nature of electrostatic interaction. For the multi-scale model of electrolytes which combines the strengths of atomistic/continuum electrolyte representations, a harmonic surface mapping algorithm is developed for fast and accurate evaluation of the electrostatic \emph{reaction potentials}. Our method reformulates the reaction potential into a sum of image charges for the near-field, and a charge density on an auxiliary spherical surface for the far-field, which can be further discretized into point charges. Fast multipole method is used to accelerate the pairwise Coulomb summation. The accuracy and efficiency of our algorithm, as well as the choice of relevant numerical parameters are demonstrated in detail. As a concrete example, for charges close to the dielectric interface, our method can improve the accuracy by two orders of magnitudes compared to the Kirkwood series expansion method.
\end{abstract}


\maketitle

\section{Introduction}

Electrostatic effect is ubiquitous in nature, and have caught broad attention in theoretical and numerical investigations, such as the criticality in electrolytes~\cite{zuckerman1997critique,romero2000coexistence, luijten2002universality}, stability of colloid suspensions~\cite{linse1999electrostatic, gutsche2007forces, dos2019like, MaPhysRevLett}, and charged biomolecular systems~\cite{HN:S:1995, lue2011macroion, SNH:COSB:2000}.
For all these studies, an accurate model of the electrolyte solvent is essential, which have aroused widespread concern up to the present~\cite{CC:JPCM:1996, BC:ARPC:2000, BH:JCC:2003, WangZMaZ2016, GJL:JSC:2016, Derbenev2016}.
The explicit solvent model~\cite{LG:ARPC:1998, Koehl:COSB:2006}, where the solvent is represented explicitly with discrete ions and water molecules, provides an accurate description of the solvent. However, its application becomes limited due to the expensive computational cost.
The implicit solvent model~\cite{RS:BC:1999, FB:COSB:2004, Baker:COSB:2005, Onufriev:ARCC:2008} replaces atomic details of the solvent with a dielectric continuum, by taking the so-called mean-field approximation of the electrolyte solvent. Such model can dramatically save the computational cost, but the detailed electrostatic interaction between water molecules/ions and the biomolecule is ignored.

An alternative that taking advantage of both models is the multi-scale hybrid model~\cite{OS:ARCC:2006,okur2006improved}. 
The hybrid model introduces a spherical cavity within which a microscopic atomistic model is used, while outside the cavity continuum theory is used to describe the (same) electrolyte solvent.
In this study, one assumes that the ionic strength lies in the weak coupling regime, where the well-known linearized Poisson--Boltzmann (LPB) equation can be used to approximate the bulk electrolyte solvent accurately~\cite{DM:CR:1990, FBM:JMB:2002, Baker:ME:2004a, ZFW:JCC:2008}.
Now the whole simulation system is splitted into two coupled atomistic/continuum regions, one further needs to decide the parameters in the hybrid model, namely the Debye length $\kappa$ and the inside/outside dielectric constants to self-consistently couple the two regions, i.e., minimize the artificial boundary effect near the spherical cavity.
For dilute electrolytes considered here, the Debye length $\kappa$ can be accurately determined as a function of the ionic densities, while the choice of inside/outside dielectric constants $\varepsilon_1$/$\varepsilon_2$ depends on different levels of microscopic descriptions inside the cavity.
There are two types of model for the inside region: i) in the hybrid $explicit/implicit$ model~\cite{OS:ARCC:2006}, both the ions and solvent molecules are treated explicitly, in that case the dielectric constant inside $\varepsilon_1$ should be taken as vacuum permittivity while the outside $\varepsilon_2$ takes the permittivity of the solvent; ii) by contrast, in the hybrid $primitive/implicit$ model~\cite{XLX:SJAM:2013} the solvent inside the cavity is modeled implicitly as a dielectric continuum but the ions are treated explicitly, in which case the inside dielectric $\varepsilon_1$ should also be taken to be that of the solvent. In recent years, the multi-scale model has been applied in Monte Carlo simulations of $1:1$ electrolytes and compared with the periodic boundary condition (PBC) using Ewald-based methods~\cite{LXX:NJP:2015, liang2017gpu}. The hybrid model shows its advantage in capturing the correct charge density profile with a smaller simulation domain, while the PBC was found to give artifacts~\cite{LXX:NJP:2015}. However, after introducing the multi-scale hybrid model, one needs to solve for the reaction potential inside the cavity due to the implicit solvent outside. Thus it becomes very important to improve the performance in solving the LPB equation in the presence of a spherical dielectric interface.

To solve the electrostatic reaction potential for the hybrid model of electrolytes, a variety of approaches have been proposed.
For water solvent, Friedman~\cite{Friedman:MP:1975} developed the image charge approximation methods, and later Abagyan and Totrov~\cite{AT:JMB:1994} proposed a modified approximation based on Friedman's approach.
These image methods have been extensively used in molecular dynamics or Monte Carlo simulations.
The multiple image charge method has also been proposed~\cite{CDJ:JCP:2007}, which can be further accelerated using the fast multipole method (FMM)~\cite{GR:JCP:1987, GR:AN:1997, CGR:JCP:1999, YBZ:JCP:2004} with $\mathcal O(N)$ complexity.  And later, the high-order accurate image charge method has been developed~\cite{DC:CCP:2007, XDC:JCP:2009, XCC:CCP:2011,XLX:SJAM:2013}. Futhermore, for an ionic solvent, Kirkwood derived the analytical solution of the reaction potential, i.e., the Kirkwood series expansion~\cite{Kirkwood:JCP:1934, TK:JACS:1957}. However, for large-scale simulations the performance of these methods is still not satisfactory, limiting the applications of the hybrid model.

In this paper, we develop a fast algorithm for the multi-scale hybrid model, which combines the image charge method for the near-field contribution, and the harmonic surface mapping algorithm (HSMA)~\cite{ZhaoLiangXuJCP2018} for the far-field.
The HSMA is a recently proposed algorithm for fast Coulomb summation as an alternative to periodic boundary condition, but its application is so far restricted to a dielectric homogeneous system. Here we extend the HSMA to the multi-scale hybrid model, which is non-trivial due to the extra dielectric interface condition and the LPB equation instead of Poisson. 
Particularly, direct-forward applying the HSMA to the hybrid model will suffer from slow convergence problem as the charges approach the interface. To overcome this issue, we further combine it with the method of images to resolve the near-field singularity. 
Numerical results demonstrate that our algorithm combining HSMA and the method of images can achieve much higher accuracy than the Kirkwood series expansion given the same truncated expansion order $p$.
We also examine the challenging case of a source charge very close to the dielectric interface, where our method improves the accuracy by two orders of magnitudes, owing to the analytical resolving of singularity with the image charges. In practice, this improvement may significantly help weaken the artificial boundary effect and reduce the size of the simulation box.

The rest of the paper is organized as follows. We describe the model and derivation of the HSMA in Sec.~\ref{sec:method}. The resulting algorithm, its computational complexity, and error analysis are summarized in Sec.~\ref{sec:algorithm}.  Then numerical results are given in Sec.~\ref{sec:results}, where the accuracy and efficiency performance are demonstrated through a few concrete examples. Finally, our conclusion and future works are summarized in Sec.~\ref{sec:conclusion}.

\section{Method}\label{sec:method}
\subsection{Model and mathematical formulations}
\begin{figure}[htbp!]
      \centering
      \includegraphics[scale=0.75]{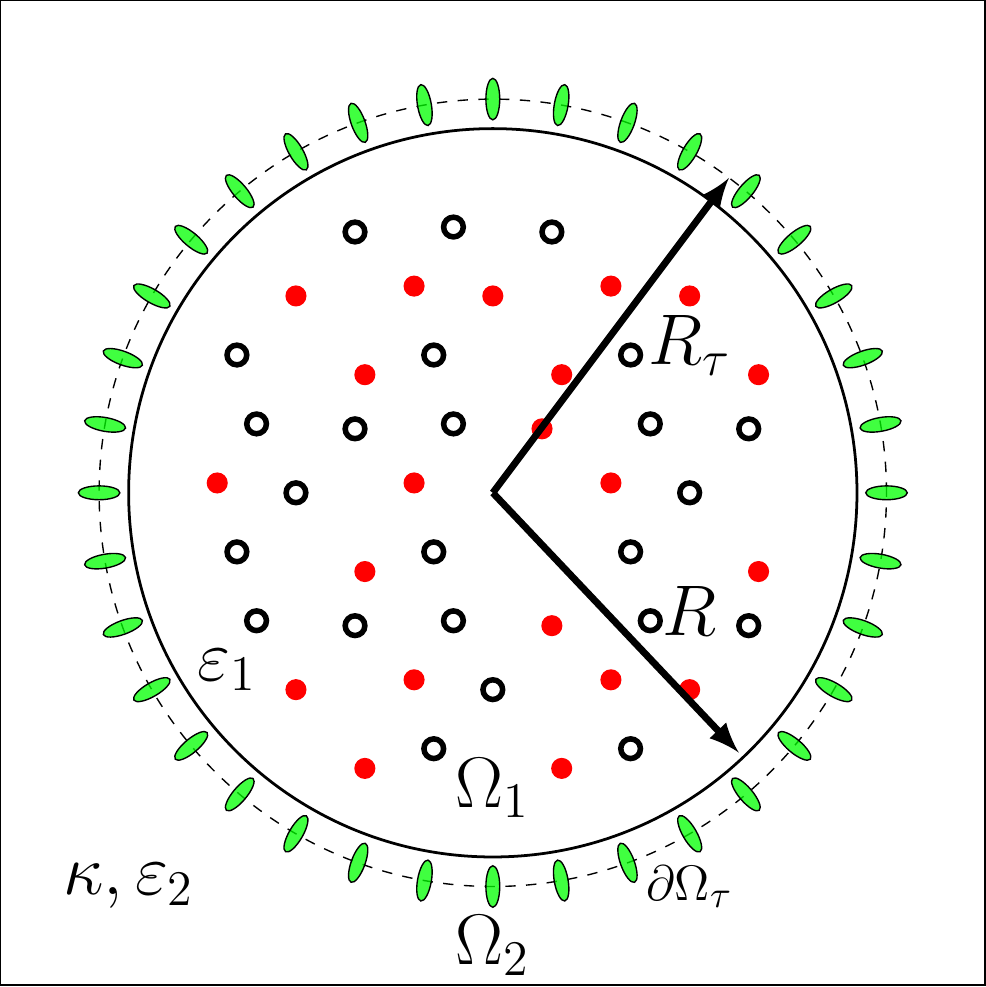}
      \caption{2D schematic illustration for a dielectric sphere immersed in an electrolyte.
$\Omega_1$ is the inside domain of the dielectric sphere, which contains explicit source charges.
$\Omega_2$ is the exterior domain, where the solvent is described as a dielectric continuum.
The dashed surface $\partial\Omega_{\tau}$ is an auxiliary surface enclosing $\Omega_1$.
In the HSMA, the reaction potential due to the continuum solvent outside $\partial\Omega_{\tau}$ is represented by point/dipole images located on it.}
      \label{fig:scheme}
\end{figure}

Consider a set of $N$ point sources located at $\bm{x}_i = (r_i, \theta_i, \phi_i )$ inside a spherical domain $\Omega_1\in\mathbb R^3$, each carrying charge $q_i$, while the outside solvent domain $\Omega_2$ is modeled as a dielectric continuum. The dielectric sphere $\Omega_1$ centered at the origin with radius $R$, as is illustrated in Fig.~\ref{fig:scheme}.
The potential inside $\Omega_1$ satisfies the Poisson equation, and the linearized Poisson-Boltzmann equation (LPB) can be used to approximate the potential in  $\Omega_2$ when the ionic strength of the electrolyte solution is week, according to the Debye-H\"uckel theory~\cite{DH:PZ:1923b}.
The electrostatic potential at $\bm{x} = (r, \theta, \phi)$ satisfies
\begin{align}
      \label{phiinsideeq}
      &-\nabla \cdot \varepsilon_1 \nabla \Phi_\mathrm{1}(\bm{x}) = 4\pi \sum_{i=1}^{N} q_i \delta(\bm{x}-\bm{x}_i), \  \textrm{in } \Omega_1,   \\
      \label{phioutsideeq}
      &-\nabla^2\Phi_2(\bm{x})+\kappa^2 \Phi_\mathrm{2}(\bm{x}) = 0, \  \textrm{in } \Omega_2,
\end{align}
where $\Phi_\mathrm{1}(\bm{x})$ and $\Phi_\mathrm{2}(\bm{x})$ are electrostatic potentials in $\Omega_1$ and $\Omega_2$, which satisfy the electrostatic interface conditions at $\partial\Omega_1$:
\begin{align}
      \label{boundarycondition}
      \Phi_\mathrm{1}=\Phi_\mathrm{2}, \ \ \varepsilon_1 \frac{\partial \Phi_\mathrm{1}}{\partial r} = \varepsilon_\mathrm{2} \frac{\partial \Phi_2}{\partial r}, \ \textrm{for} ~~\bm x\in\partial\Omega_1.
\end{align}
Note that $\varepsilon_1$ and $\varepsilon_2$ are dielectric constants in $\Omega_1$ and $\Omega_2$, respectively, $\varepsilon_1=\varepsilon_2$ if primitive model is used in $\Omega_1$, and $\varepsilon_1\approx 1$ if all the molecules are treated explicitly in $\Omega_1$.
 $\kappa$ is the inverse Debye length, $\kappa =(4\pi l_\mathrm{B} \sum_{j} \lambda_{j}z_j^2)^{1/2} $, where index $j$ runs over all the ion species, and $\lambda_{j}$ and $z_j$ are the bulk concentration and valence of the $j$th ion species.
 $l_\mathrm{B}$ is the solvent Bjerrum length, which equals $7.14$\AA~for water at room temperature.
Finally, the far-field boundary condition is $\Phi_\mathrm{2}\to 0$ as $r\to \infty$.

Note that in solving the electrostatics for an arbitrary shaped dielectric interface, boundary integral equation methods have been developed~\cite{GK:JCP:2013, bardhan2009discretization,berti2012comparison}. However, for spherical geometry, more efficient methods can be developed in solving the LPB as was summarized in the introduction, or even for the nonlinear PB equation~\cite{vishnyakov2007nonlinear}.  
One may argue that the spherical geometry as depicted in Fig.~\ref{fig:scheme} is very specialized, has rather limited applications. However, in the multi-scale modeling of electrolytes, the spherical interface is artificially introduced for its simplicity, and serves as an alternative to the popular periodic boundary condition. 

\subsection{Kirkwood series expansion revisited}
We first revisit the Kirkwood series expansion~\cite{Kirkwood:JCP:1934, TK:JACS:1957} for solving Eqs.~\eqref{phiinsideeq}--\eqref{boundarycondition}.
The potential $\Phi_1$ inside $\Omega_1$ can be written as the sum of two contributions,
\begin{align}
      \label{phiin}
      \Phi_\mathrm{1}(\bm{x})= \Phi_{\mathrm{Coul}}(\bm{x}) + \Phi_{\mathrm{RF}}(\bm{x}),
\end{align}
where $\Phi_{\mathrm{Coul}}$ is the Coulomb potential due to the point source charges,
\begin{align}
      \label{phicoul}
      \Phi_\mathrm{Coul}(\bm{x})= \sum_{i=1}^{N}\frac{q_i}{4\pi\varepsilon_1 |\bm{x} - \bm{x}_i |}.
\end{align}
One can further expands the Coulomb potential using spherical harmonics~\cite{Jackson::2001}, obtaining the following expression for $\Phi_{\mathrm{Coul}}$:
\begin{align}
      \label{phicoul_final}
      \Phi_\mathrm{Coul}(\bm{x})  =  \sum_{i=1}^{N}\sum_{n,m}^{\infty}\frac{q_i}{(2n+1)\varepsilon_1} \frac{r_{<}^{n}}{r_{>}^{n+1}}  Y_n^{m *}(\theta_i, \phi_i)Y_n^m(\theta, \phi),
\end{align}
where $\sum_{n, m}^{\infty}$,is short for the multipole expansion summation $\sum_{n=0}^{\infty}\sum_{m=-n}^{n}$, $r_{<} (r_{>})$ is the smaller (larger) value between $r_i$ and $r$, and $Y_n^m(\theta, \phi)$ is the spherical harmonic function of degree $n$ and order $m$. Note that the superscript * denotes complex conjugate.
The second contribution $\Phi_{\mathrm{RF}}$ is the reaction potential due to the exterior continuum solvent.
Since $\Phi_{\mathrm{RF}}$ is a harmonic function, it can also be expanded in terms of spherical harmonics,
\begin{align}
      \label{phirf}
      \Phi_\mathrm{RF}(\bm{x})=\sum_{n, m}^{\infty}A_n^m r^n Y_n^m(\theta, \phi),
\end{align}
where $A_n^m$ are the undetermined expansion coefficients.

Analogously, the potential $\Phi_2$ in the exterior region $\Omega_2$ can also be expanded as,
\begin{align}
      \label{phiex}
      \Phi_\mathrm{2}(\bm{x}) = \sum_{n, m}^{\infty} B_{n}^m k_n(\kappa r)Y_n^m(\theta, \phi),
\end{align}
where $B_n^m$ are unknown coefficients and $k_n(\cdot)$ is the modified spherical Hankel function of order $n$, defined as~\cite{AS::1964}:
\begin{equation}
      k_n(\mu)=\frac{\pi e^{-\mu}}{2\mu}\sum_{l=0}^{n}\frac{(n+l)!}{l!(n-l)!}\frac{1}{(2\mu)^l}.
\end{equation}

Now since both $\Phi_1$ and $\Phi_2$ are expanded using spherical harmonics, one can further substitute Eqs.~\eqref{phiin}--\eqref{phiex} into the interface conditions~\eqref{boundarycondition} to solve for the unknown coefficients $A_n^m$ and $B_n^m $.
By the orthogonality of the spherical harmonics, one obtains the following expressions for $A_n^m$ and $B_n^m$,
\begin{align}
      \label{Anm}
      A_n^m(\mu) &= \frac{\varepsilon(n+1)S_n(\mu) + 1}{\varepsilon n S_n(\mu) - 1}  
        \sum_{i=1}^{N}\frac{q_i }{(2n+1)\varepsilon_1 R }\frac{Y_{n}^{m*}(\theta_i, \phi_i) }{r_{\mathrm K, i}^n }, \\
      \label{Bnm}
      B_n^m(\mu) &= \frac{\varepsilon(2n+1)}{\varepsilon n k_n(\mu) - \mu k_n^{'}(\mu) } 
       \sum_{i=1}^{N}\frac{q_i }{(2n+1)\varepsilon_1 R} \left( \frac{r_i}{R} \right)^n Y_{n}^{m*}(\theta_i, \phi_i),
\end{align}
where  $\bm{r}_{\mathrm K, i} $ is the so-called Kelvin image point, defined as $\bm{r}_{\mathrm K, i}= (R/r_i)^2\bm{x}_i$,
and $\varepsilon = \varepsilon_1/\varepsilon_2$, $ \mu=\kappa R$ and $S_n(\mu)=\frac{k_n(\mu)}{\mu k_n'(\mu)}$.

Finally, it is worth noting that $S_n(\mu)$ has the following asymptotic approximations, as $\mu\to\infty$~\cite{XDC:JCP:2009,XC:SR:2011},
\begin{align}
      \label{Snuinfty}
      S_n(\mu) = -\frac{1}{n+1+\mu} + \mathcal O \left(\frac{1}{\mu^2} \right), 
\end{align}
and as $\mu\to0$,
\begin{align}
      \label{Snuzero}
      S_n(\mu) = -\frac{1}{n+1+\mu} + \mathcal O \left(\mu \right), 
\end{align}
thus the formula gives correct leading-order asymptotics for both low and high concentrations of the electrolytes.
With the asymptotic formulas, the reaction potential can be further simplified into an image charge expression, which will be described in Sec.~\ref{sec:image}.

\subsection{Image charge representation}\label{sec:image}
In this section, one derives an image charge representation for the reaction potential.
First, consider the expansion coefficients of $A_n^m(\mu)$ in Eq.~\eqref{Anm}, denoted here as $M_n(\mu)$:
  \begin{align}
            \label{Anuapp1}
           M_n(\mu) = \frac{\varepsilon(n+1)S_n(\mu) + 1}{\varepsilon n S_n(\mu) - 1}.
      \end{align}
By substituting the asymptotic formula (i.e. Eq.~\eqref{Snuinfty}) into Eq.~\eqref{Anuapp1}, $M_n(\mu)$ can be decomposed into three parts:
      \begin{align}
            \label{Anuapp}
           M_n(\mu)=\gamma + \frac{\bar{\delta} }{n+\sigma} +\widehat{M}_n(\mu),
      \end{align}
where
\begin{align}
& \gamma=\frac{\varepsilon -1 }{\varepsilon +1},~~
\sigma = \frac{1+\mu}{1+\varepsilon}, ~~ \bar{\delta} = \gamma(1-\sigma)-\frac{\mu}{1+\varepsilon}.
\end{align}
Note that $\widehat{M}_n(\mu)$ denotes the $\mathcal O(1/\mu^2)$ higher-order terms in the asymptotic expansion~\eqref{Snuinfty}. We do not attempt to find the explicit expression of $\widehat{M}_n(\mu)$, its contribution will be mapped onto an auxiliary surface in the HSMA method.

Next, substitute Eq.~\eqref{Anuapp} into Eqs.~\eqref{phirf} and~\eqref{Anm}, and further apply the following identity,
    \begin{align}
        \frac{1}{n+\sigma} = r_{\mathrm K, i}^{n+\sigma}\int_{r_{\mathrm K, i}}^{\infty}\frac{1}{x^{n+\sigma+1}}dx,\end{align}
which is valid for all $n \geqslant 0 $ with $\sigma > 0$ is a constant.
The following expression is obtained for the reaction potential $\Phi_{\mathrm{RF}}$ as the sum of Kelvin images, line image densities, and a spherical harmonic expansion which can be regarded as a higher-order correction term:
      \begin{align}
            \label{Phihybrid}
             \Phi_{\mathrm{RF}}(\bm{x}) &= \underbrace{\sum_{i=1}^{N} \frac{q_i \gamma r_{\mathrm K,i} }{4\pi\varepsilon_1R|\bm{x} - \bm{r}_{\mathrm K,i}|} }_{\text{Kelvin images}} + \underbrace{ \sum_{i=1}^{N}  \int_{r_{\mathrm K,i}}^{\infty} \frac{ q_i \bar{\delta} \left(t/r_{\mathrm K,i}\right)^{-\sigma}}{4\pi\varepsilon_1 R |\bm{x} - \bm{t} |} dt}_{\text{Line image densities}} \nonumber\\
            &+ \underbrace{\sum_{i=1}^N\sum_{n, m}^{\infty}  \frac{q_i  \widehat{M}_n(\mu)}{(2n+1) \varepsilon_1 R} \left(\frac{r}{r_{\mathrm K,i} } \right)^n Y_n^{m*}(\theta_i, \phi_i) Y_n^m(\theta, \phi)}_{\text{Higher-order correction term}}.
      \end{align}
If $\varepsilon = 1$ (i.e., no dielectric jump at the interface), the Kelvin images will vanish ($\gamma=0$), but the line images and the higher-order correction term will still be non-zero, due to the jump in the inverse Debye length $\kappa$ at the interface.

In~\cite{XDC:JCP:2009}, numerical results show that keeping the first two leading-order terms can approximate the reaction potential accurately  when the source charges are \emph{not} very close to the dielectric interface.
In this work, one aims to keep the third correction term, which will allow us to obtain high-order accuracy even when source charges are close to the interface. And since the direct calculation of the correction term is time consuming, one introduces the harmonic surface mapping technique below to simplify it as a sum of images further.

\subsection{Harmonic surface mapping}
The harmonic surface mapping algorithm is a recently proposed fast algorithm~\cite{ZhaoLiangXuJCP2018} for solving the Poisson equation with either periodic/non-periodic boundary conditions. But it has not yet been applied to such hybrid model, where one needs to solve the LPB equation, and there exists a spherical dielectric interface.

Let us introduce the auxiliary spherical surface $\partial\Omega_{\tau}$. As was shown in Fig.~\ref{fig:scheme},
it is concentric with $\partial \Omega_1$ and encloses the whole interior domain $\Omega_1$.
The radius of the auxiliary surface is $R_{\tau}=(1+\tau)R$, note that $R$ is the radius of $\Omega_1$ and we have an adjustable parameter $\tau > 0$ (practically setting the value of $\tau$ will be discussed in Sec.~\ref{sec:results}).
Then the line image integrals in Eq.~\eqref{Phihybrid} can be divided as $\int_{r_{\mathrm K,i} }^{\infty} = \int_{r_{\mathrm K,i} }^{R_{\tau}} + \int_{R_{\tau}}^{\infty}$. 
One uses the trapezoidal rule to approximate the first line integral on $[r_{\mathrm K,i}, R_{\tau}]$,
then the quadrature weight assigned at the Kelvin point $\bm{r}_{\mathrm K,i}$ is $ \bar{\delta} \frac{(R_{\tau} - r_{\mathrm K,i} )}{2}\frac{q_i}{R}$. Note that its location overlaps with the original Kelvin image, so one just modifies the original Kelvin image and obtains the new Kelvin image magnitude
\begin{align}
q_{\mathrm K,i}= \frac{q_i}{R} \left[\gamma r_{\mathrm K,i} + \bar{\delta} \frac{(R_{\tau} - r_{\mathrm K,i} )}{2} \right].
\end{align}
The other trapezoidal point at $\bm{R}_{\tau}$ and the numerical discretization error can be both absorbed into the correction term, thus the modified harmonic coefficient $\widehat{M}_n^{'}(\mu)$ is defined,
  \begin{align}
            \widehat{M}_n^{'}(\mu) = M_n(\mu)- \gamma - \bar{\delta} \frac{(R_{\tau} - r_{\mathrm K, i} )}{2 r_{\mathrm K, i}}.
      \end{align}
Then the reaction potential $\Phi_{\mathrm{RF}}$ can be rewritten as,
      \begin{align}
            \label{PhihybridApprk}
           \Phi_{\mathrm{RF}}(\bm{x}) = \sum_{i=1}^{N} \frac{q_{\mathrm K,i} }{4\pi\varepsilon_1|\bm{x} - \bm{r}_{\mathrm K, i} |}  
             + \sum_{i=1}^N\sum_{n, m}^{\infty}  \frac{q_i \widehat{M}_n^{'}(\mu) }{(2n+1) \varepsilon_1 R} \left(\frac{r}{r_{\mathrm K,i} } \right)^n Y_{n}^{m*}(\theta_i, \phi_i) Y_n^m(\theta, \phi).
      \end{align}
Note that the trapezoidal rule is used to obtain the modified Kelvin image magnitude $q_{\mathrm K,i}$, but the above expression for $\Phi_{\mathrm{RF}}$ is still exact.
However, directly calculating the second correction term in Eq.~\eqref{PhihybridApprk} will be again time-consuming, so one should discuss below how to handle it computationally through the HSMA approach.

We first define the correction term in Eq.~\eqref{PhihybridApprk} (truncated at order $p$) as
\begin{align}
\label{phim}
      \Phi_{\mathrm{corr}}(\bm{x})\approx\sum_{n, m}^{p} \widehat{A}_n^m r^n Y_n^m(\theta, \phi),
\end{align}
where
\begin{align}
 \widehat{A}_n^m = \sum_{i=1}^N \frac{q_i \widehat{M}_n^{'}(\mu) }{(2n+1) \varepsilon_1 R} \frac{ Y_{n}^{m*}(\theta_i, \phi_i) }{r_{\mathrm K,i}^n }.
 \end{align}
Through the Green's second identity and the fact that $\Phi_{\mathrm{corr}}$ is a harmonic function, one can convert $\Phi_{\mathrm{corr}}$ into a surface integral on the auxiliary surface $\partial \Omega_{\tau}$,
\begin{align}
      \label{GreenSecond}
      \Phi_{\mathrm{corr}}(\bm{x}) = \oiint_{\partial \Omega_{\tau}}\left[G (\bm{x}, \bm{y})\frac{\partial \Phi_{\mathrm{corr}}(\bm{y} )}{\partial \bm{\nu}_y} 
       - \Phi_{\mathrm{corr}}(\bm{y} )\frac{\partial G(\bm{x}, \bm{y})}{ \partial \bm{\nu}_y}  \right]dS_y,
\end{align}
where $G(\bm{x}, \bm{y})=\frac{1}{4\pi|\bm{x}-\bm{y}|}$ is the Green's function for Poisson equation in free space and $\bm{\nu}_y$ is the unit outward normal vector at $\bm{y}$.
One further uses the central difference scheme to approximate $\frac{\partial G}{ \partial \bm{\nu}_y}$ in Eq.~\eqref{GreenSecond}, i.e.,
\begin{align}
      \label{approxG}
      \frac{\partial G}{ \partial \bm{\nu}_y} \approx \frac{1}{\Delta y}[G(\bm{x}, \bm{y}^{+} )-G(\bm{x}, \bm{y}^{-})],
\end{align}
where $\bm{y}^{\pm} = (y \pm \Delta y/2, \theta, \phi)$ and $\Delta y$ is the central difference step size.
Then $\Phi_{\mathrm{corr}}(\bm{x})$ can be expressed as the sum of three surface integrals,
      \begin{align}
      \label{threeintegrals}
             \Phi_{\mathrm{corr}}(\bm{x})  \approx \oiint_{\partial \Omega_{\tau}}  \frac{\partial_{\bm{y}}\Phi_{\mathrm{corr}}(\bm{y}) }{4\pi|\bm{x}-\bm{y}|}dS_y 
             + \oiint_{\partial \Omega_{\tau}^-}\frac{\Phi_{\mathrm{corr}}(\bm{y}) /\triangle y }{4\pi |\bm{x}-\bm{y}^{-} |}  dS_y
             -  \oiint_{\partial \Omega_{\tau}^+}\frac{\Phi_{\mathrm{corr}}(\bm{y}) / \triangle y}{4\pi|\bm{x}-\bm{y}^{+} |}dS_y.
      \end{align}
The first term represents a surface charge density, while the second and third terms are essential a central difference approximation for a surface dipole density.
It is worth noting that since $\bm x\in\Omega_1$ while $\bm y\in\partial\Omega_\tau$, all three integrands in Eq.~\eqref{threeintegrals} are non-singular.
Thus the Fibonacci numerical integration scheme~\cite{hannay2004fibonacci} can be applied, which achieves $\mathcal O( N_{\tau}^{-6})$ convergence for $N_{\tau}$ discretized grid points.

Suppose $f(\bm{y})$ is a non-singular integrand, the Fibonacci integration method discretizes the surface integral over $f(\bm{y})$ on a sphere $\partial\Omega_\tau$ as
      \begin{align}
            \oiint_{\partial \Omega_{\tau}}f(\bm{y})dS_y \approx
            \frac{2\pi R_{\tau}^2}{F_2}\sum_{j=0}^{F_2}[1+\cos(\pi z_j)] 
            \times [f(\bm{y}_{2j+1})+f(\bm{y}_{2j+2})],
      \end{align}
      where $z_j=(-1+2j/F_2)$, $\bm{y}_{2j+1}=(R_{\tau}, \arccos(z_j+\sin(\pi z_j)/\pi), \pi jF_1/F_2)$, $\bm{y}_{2j+2}=(R_{\tau}, \arccos(z_j+\sin(\pi z_j)/\pi), \pi+\pi jF_1/F_2)$, and $F_1 < F_2$ are two successive Fibonacci numbers.
Therefore, after discretization using the Fibonacci integration scheme, Eq.~\eqref{threeintegrals} can be approximated by a sum of discrete images located on $\Omega_{\tau}$ and $\Omega_{\tau}^\pm$, i.e.,
      \begin{align}
            \label{phicorrfinall}
            \Phi_{\mathrm{corr} }(\bm{x}) \approx \sum_{j=1}^{N_{\tau}}\frac{\bar{q}_j}{|\bm{\bar{y}}_j - \bm{x}|},
      \end{align}
      where $N_{\tau} = 6(F_2 + 1) $ is the total number of images, and $\bar{q}_j $  and $ \bm{\bar{y}}_j$ are the charge and location from the  Fibonacci numerical integration.
It is a noteworthy fact that  $N_\tau$ is independent of the total number of sources $N$.
Finally, one substitutes Eq.~\eqref{phicorrfinall} to the reaction potential Eq.~\eqref{PhihybridApprk} and combines it with the Kelvin images, then obtains a simple expression for $\Phi_{\mathrm{RF}}$ as a Coulomb sum of $N + N_{\tau}$ image charges, i.e.,
      \begin{align}
            \label{phirffinall}
            \Phi_{\mathrm{RF} }(\bm{x}) \approx \sum_{j=1}^{N + N_{\tau} }\frac{Q_j}{|\bm{Y}_j - \bm{x}|},
      \end{align}
where
$Q_j = q_{\mathrm K,j}, \bm{Y}_j = \bm{r}_{\mathrm K,j}$ for $j = 1, \dots, N$,
and $Q_j = \bar{q}_j, \bm{Y}_j = \bm{\bar{y}}_j$ for $j = N+1, \dots, N + N_{\tau}$.
For the case of $\varepsilon = 1$, although the original Kelvin images vanish, the total number of the image charges of the reaction potential is still $N + N_{\tau}$ due to the trapezoidal rule used to discretize the line images.

\section{Algorithm, complexity and error analysis}\label{sec:algorithm}
We now describe the algorithm steps and its computational complexity, (as summarized in Algorithm~\ref{alg:hsma}).

\begin{algorithm}
      \caption{Harmonic surface mapping algorithm}
      \label{alg:hsma}
      \begin{algorithmic}[1]
      \REQUIRE  Spherical harmonic expansion truncated order $p$, dielectric constants $\varepsilon_{1,2}$ and inverse Debye length $\kappa$ for the electrolyte, dielectric sphere radius $R$, auxiliary surface radius $R_{\tau} = (1 + \tau)R$, source charge locations $r_i$ and charge magnitudes $q_i$ for $i = 1,\dots,N$.
        \STATE Construct $2p$ Gauss-Legendre quadrature nodes along $\theta$ direction  and $2p$ equi-spaced weights along $\phi$ direction on $\partial \Omega_1$, which will be used for the spherical harmonic transform.
      \STATE Use FMM to calculate the potential generated by the source charges at the quadrature nodes on $\partial \Omega_1$. This step has complexity $\mathcal O(N+p^2)$.
     \STATE  Discrete spherical harmonic transformation is performed to obtain the spherical harmonic expansion coefficients $\widehat{A}_n^m(\mu)$. This step requires $\mathcal O(Np^3)$ operations.
      \STATE Generate the Kelvin image charges $q_K$, $\bm{r}_{\mathrm K,i}$, $i = 1, \dots, N$ inside $\Omega_{\tau}$ and the Fibonacci integration points and weights $Q_j$, ${\bm{R}_j, j=1, \dots, N_{\tau}}$ on $\partial \Omega_{\tau}$. This step has complexity $\mathcal O(N + N_{\tau}p^2)$.
      \STATE Use FMM to evaluate the electrostatic potential/field at source charge locations. This step costs $\mathcal O(N + N_{\tau})$.
      \end{algorithmic}
\end{algorithm}

The numerical error of HSMA comes from three parts:
(a) The $p$-th order truncation error of the spherical harmonic expansion in Eq.~\eqref{PhihybridApprk}.
An error estimation for this part was given in~\cite{GUMEROV2014307}, i.e.,
if truncated at order $p$, the truncation error $ \mathcal E_{\mathrm{trunc}} \sim \mathcal O\left(\frac{1}{1+\tau} \right)^p$;
(b) The discretization error from the central difference in Eq.~\eqref{approxG} for calculating $\partial G/\partial \nu_y$, $\mathcal E_{\mathrm{diff}}\sim\mathcal O(\Delta y^2)$;
(c) The Fibonacci numerical integration error, $\mathcal E_{\mathrm{Fibo}}\sim\mathcal O(N_{\tau}^{-6})$.
In practice, one finds that the truncation error from part (a) is the dominant part, as long as a reasonable $\Delta y$ and $N_{\tau}$ is chosen, the errors from parts (b) and (c) are minor.
However, it should be noted that given the same truncation order $p$, the HSMA can achieve better accuracy than merely using the Kirkwood series, due to the fact that the numerical singularity is mainly caused by the Kelvin images, which has been handled here analytically. Numerical evidence will be shown in Sec.~\ref{sec:results}.

\section{Numerical results}\label{sec:results}
In this section, one tests the performance of the HSMA in terms of both accuracy and efficiency.
In all the calculations, one fixes $R=1, \mu=5$, $\varepsilon_1=2$ and $\varepsilon_2=80$, and varies parameters $\tau$ and $p$ to check the accuracy.
Since the error from the central difference is minor, one takes the step size to be $\Delta y = 10^{-5}R_{\tau}$.
And one takes the results from the Kirkwood series truncated at sufficiently large $p$ (for the worst case here one takes $p=201$) as the reference solution.

\subsection{Accuracy tests}\label{sec:accuracy}
One first tests the accuracy of the HSMA by considering a unit source charge inside the spherical dielectric interface.
Suppose a unit point source located at $\bm{x}_s=(r_s, \theta_s, \phi_s)$ inside $\Omega_1$, i.e. $r_s \in (0,1)$, one considers the error in its self energy, the self energy $E_{\mathrm{self}}$ is defined as
\begin{align}
E_{\mathrm{self}}=\frac{1}{2}\Phi_{\mathrm{RF}}(\bm{x}_s, \bm{x}_s).
\end{align}

\begin{figure}[htbp!]
      \centering
      \includegraphics[scale=0.35]{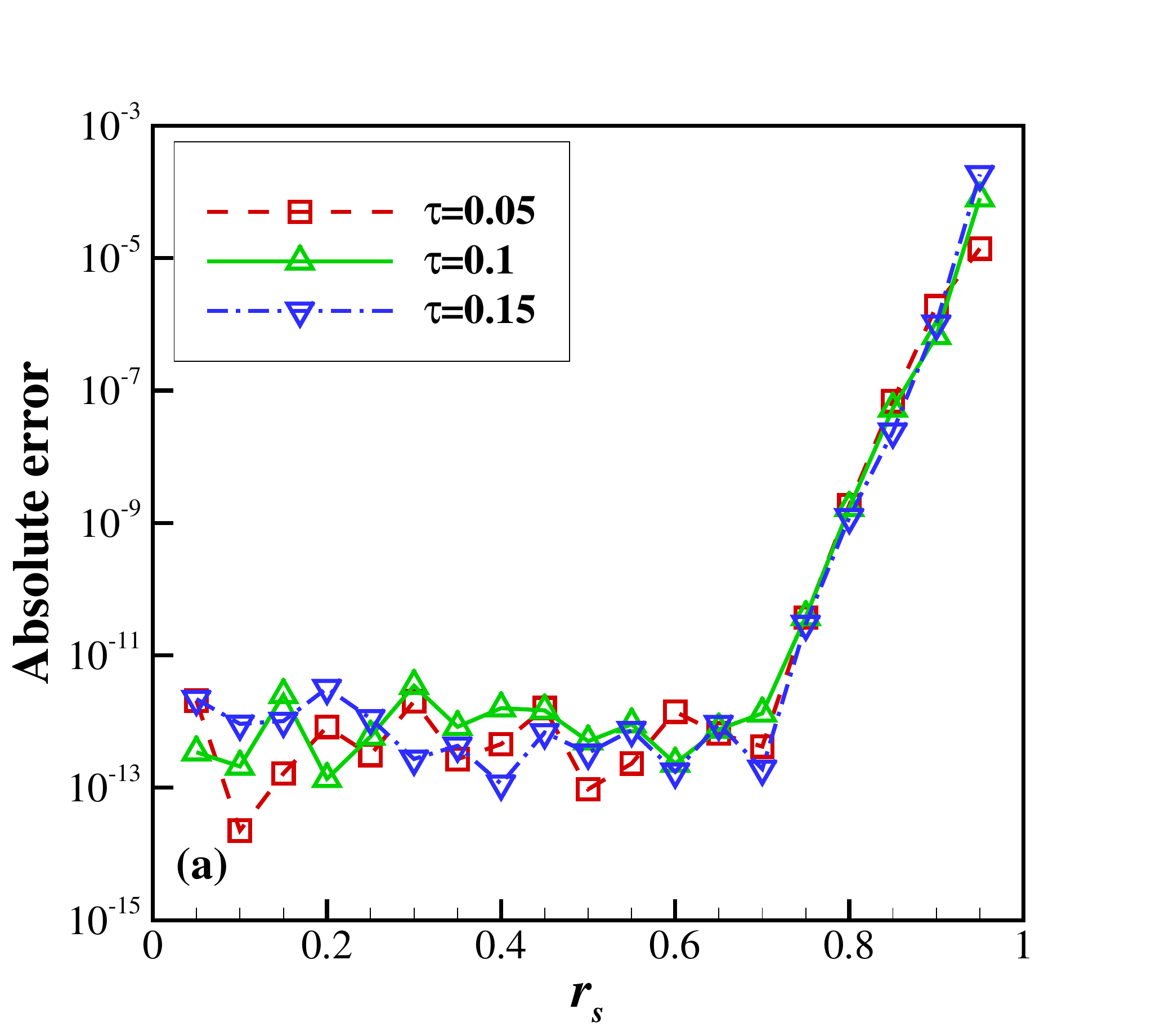}
      \includegraphics[scale=0.35]{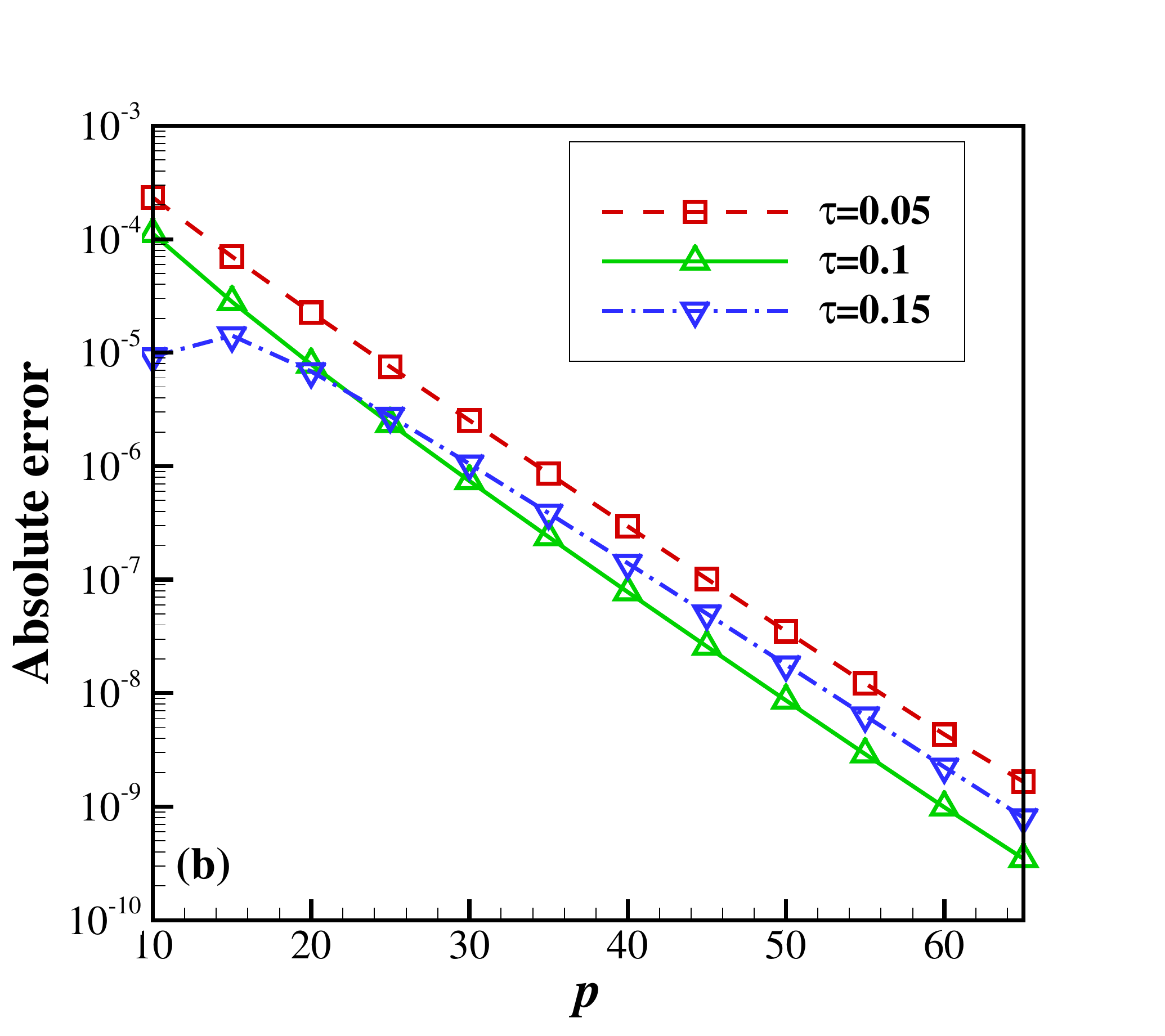}
      \caption{(a) Absolute errors of the self energy as a function of source charge position $r_s$ for $\tau=0.05$, $0.1$ and $0.15$, one fixes $p=30$ and $F_2=1597$. (b) Absolute errors of the self energy as a function of truncated order $p$ for $\tau=0.05$, $0.1$ and $0.15$ with $r_s=0.9$, one fixes $F_2=6765$.}
      \label{fig:AbsEtaudepend}
\end{figure}

One first tests the error dependence on the adjustable parameter $\tau$. In Fig.~\ref{fig:AbsEtaudepend} (a), one shows the numerical error in $E_{\mathrm{self}}$ as a function of the source charge location $r_s$ with $\tau=0.05$, $0.1$ and $0.15$, while fixing $p=30$ and $F_2=1597$.
One observes that for $r_s$ ranging from $0$ to $0.95$, the error is not very sensitive about the value of $\tau$. As expected, one sees that the error increases a few orders of magnitudes as the charge approaches the interface. But even when $r_s=0.95$, the HSMA can still obtain absolute error~$\sim 10^{-4}$.
In Fig.~\ref{fig:AbsEtaudepend} (b), one also tests the error in $E_{\mathrm{self}}$ as a function of the truncated order $p$ for the same set of values for $\tau$. Consistently, one finds that for $p$ ranging from $10$ to $65$, the error does not change much for different $\tau$ values. And as expected, the error decays exponentially as a function of $p$, e.g., the method exhibits spectral convergence in $p$. As a result, in the following numerical tests, one will fix $\tau=0.1$.

\begin{figure}[htbp!]
      \centering
      \includegraphics[scale=0.4]{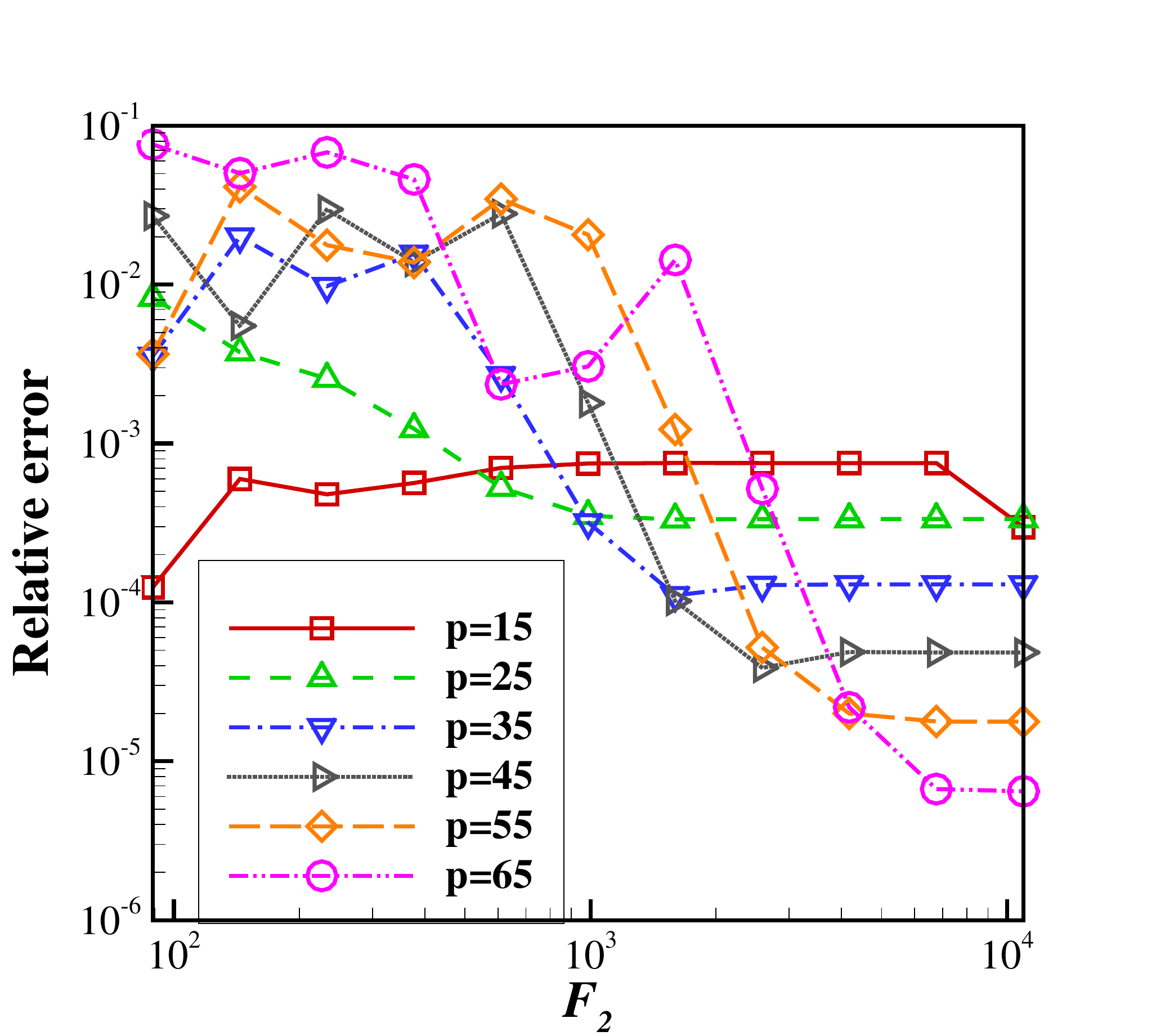}
      \caption{Relative error in the self energy as a function of  the Fibonacci number $F_2$, with truncation order $p$ ranging from 15 to 65. one fixes the charge location to be $r_s=0.95$ and $\tau=0.1$. }
      \label{fig:AbsloteError}
\end{figure}

\begin{figure}[htbp!]
      \centering
       \includegraphics[scale=0.4]{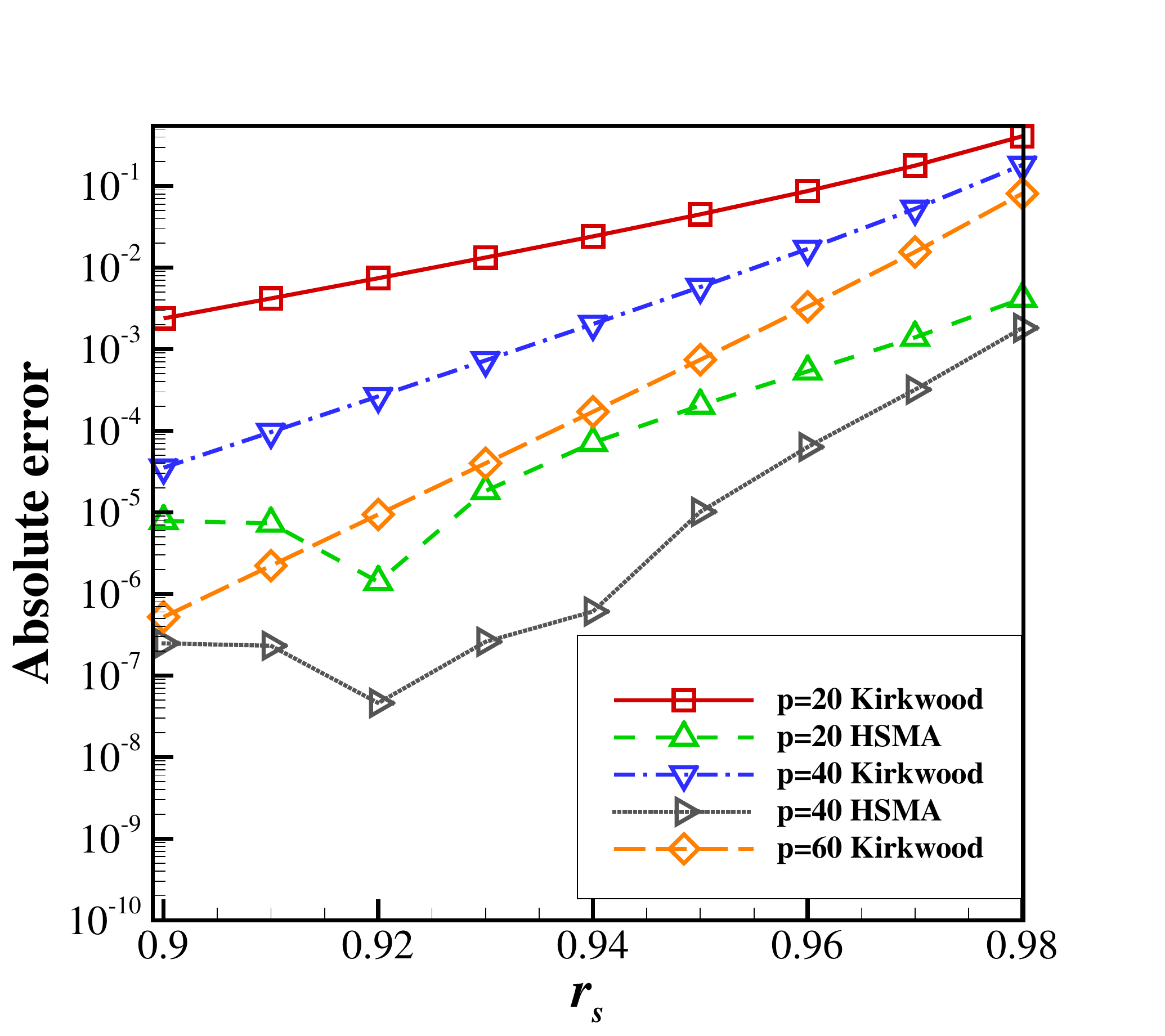}
      \caption{ Absolute errors in the self energy as a function of the source charge location $r_s\in [0.9, 0.98]$ with different values of $p$, results from both HSMA and Kirkwood series expansion are shown for comparison. Here one fixes $F_2=1597$.}
      \label{fig:hsmaKirkwoodcompare}
\end{figure}

One now moves to the error dependence on the truncation order $p$ and the Fibonacci number $F_2$ used in the numerical integration. Here one will focus on a challenging case by taking $r_s=0.95$, i.e., the charge is very close to the interface.
The results are shown in Fig.~\ref{fig:AbsloteError}.
First, one observes that for different values of $p$, the error converges very soon as $F_2$ increases due to the high-order convergence of the Fibonacci integration scheme, e.g., for the case $p=65$, the error saturates if one takes $F_2\geq 6765$.
Second, as long as one uses sufficiently large value for $F_2$, the magnitude of the saturated error is decided by the truncation order $p$ one chooses. For the case $r_s=0.95$, if one wants to achieve 5-digits accuracy, then one needs to choose $p\sim 55$.
Finally, one also compares the HSMA with the original Kirkwood series solution with the same truncation order $p$.
As is shown in Fig.~\ref{fig:hsmaKirkwoodcompare}, one observes that HSMA can improve the accuracy by two orders of magnitudes, compare to the Kirkwood series expansion with the same truncation order $p$.
For example, HSMA with $p=20$ achieves even better accuracy than the Kirkwood series with $p=40$ over the whole range of $r_s\in [0.9, 0.98]$.
Moreover, by choosing $p=20$ for HSMA, it is guaranteed that the absolute error remains less than $10^{-2}$, while the Kirkwood series can not achieve the same goal even if one takes $p=60$.
Thus the HSMA has a clear advantage in terms of accuracy compared with the Kirkwood series approach.

\subsection{CPU time tests}
\begin{figure}[htbp!]
      \centering
      \includegraphics[scale=0.4]{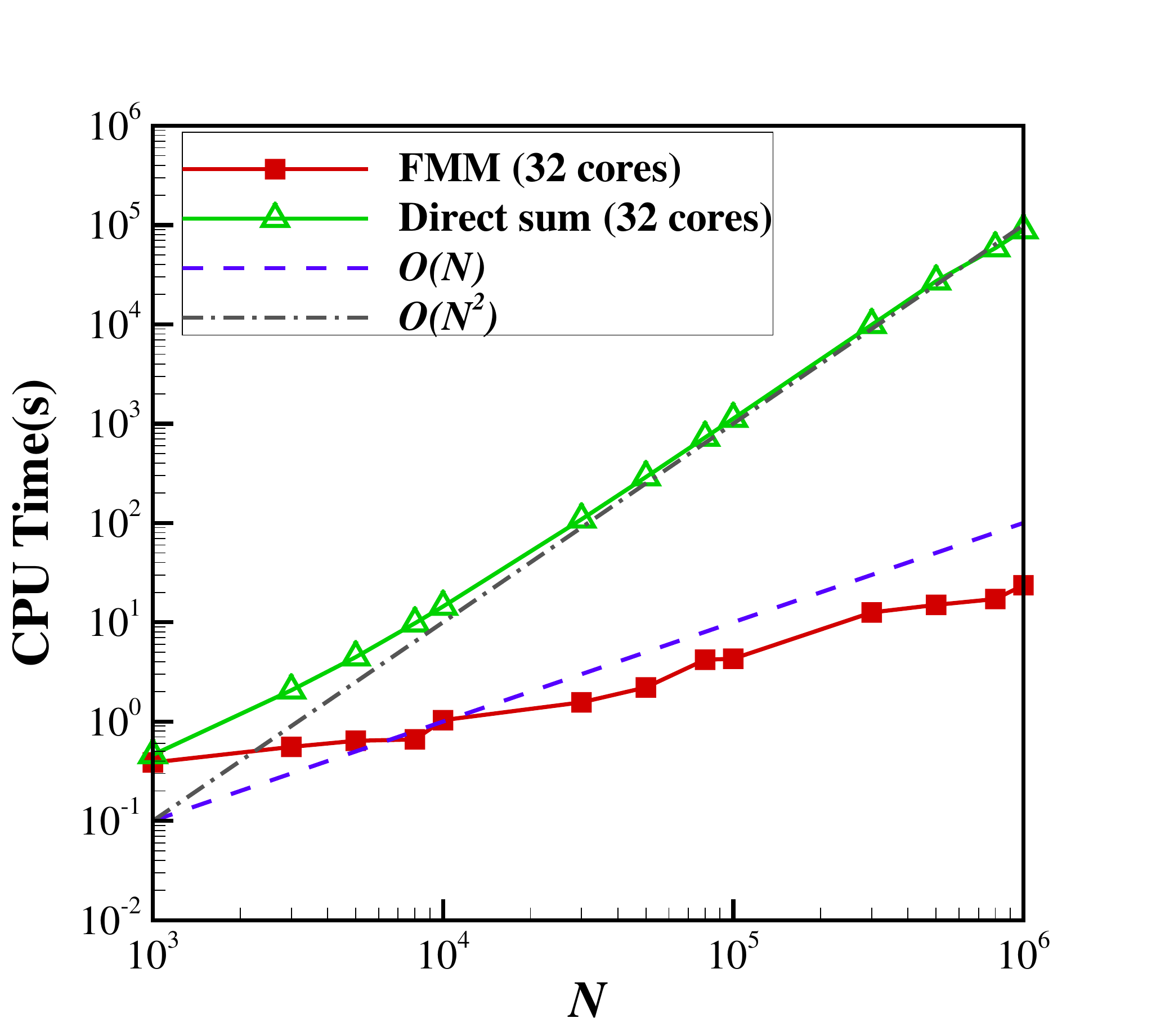}
      \caption{CPU time for calculating the reaction potential of $N$ random generated point charges, with $N$ ranging from $10^3$ to $10^6$. The results of HSMA both with and without FMM acceleration are shown here for comparison.
The parameters of HSMA chosen here are $\tau=0.1, P=20, F_2=1597$.}
      \label{fig:rs_AbsloteErrorTime}
\end{figure}

Here one tests the CPU time performance of our method for a large number of source charges inside the dielectric sphere.
One uses FMM (the software package FMM3DLIB~\cite{FMMLIB3D}) to accelerate the pairwise Coulomb summations, with the FMM precision fixed to be $10^{-6}$.
The timing results are obtained on a 64-core workstation(4 AMD operation Processors Model 6272, 2.1 GHz with 16 cores each), and one uses 32 cores for each run.
The parameters of the HSMA are chosen to be $\tau=0.1, p=20, F_1=987$ and $F_2 = 1597$.
As was tested in Sec.~\ref{sec:accuracy}, the parameters chosen here are sufficient to obtain numerical error less than $10^{-5}$ if $r_s < 0.93R$.
In the numerical tests here, one randomly generates $N$ source charges inside the dielectric sphere, with $N$ ranging from $10^3$ to $10^6$, and one calculates the reaction potential of the system.
As is shown in Fig.~\ref{fig:rs_AbsloteErrorTime}, one finds that HSMA accelerated by FMM can achieve linear $\mathcal O(N)$ scaling. And compare with the direct sum, the break-even is around $N=1000$.
For large-scale simulations, say if the system contains $10^6$ source charges, the CPU time of the HSMA accelerated by FMM is $24$s, while with direct sum the cost becomes $8.9 \times 10^4$s.
Thus the FMM-accelerated HSMA can be very attractive for large-scale simulations of charged particles immersed in electrolytes.

\section{Conclusion}\label{sec:conclusion}
We have developed a harmonic surface mapping algorithm for calculating the electrostatic reaction potentials in multi-scale model of electrolytes. 
Based on the Kirkwood series solution, the asymptotic expansion is first used to rewrite the reaction potential into the sum of Kelvin images, line images, and an extra correction term.
Then an auxiliary surface is introduced, by using the Green's second identity, allowing us to transform the correction term into an integral on the auxiliary surface.
Further combined with the Fibonacci integration scheme, we manage to rewrite the reaction potential into a simple pairwise Coulomb summation, which can be further accelerated by FMM in large-scale simulations to achieve linear scaling.
Numerical tests demonstrate that HSMA can achieve much better accuracy comparing with the Kirkwood series solution. Particularly, even when a source charge is very close to the dielectric interface, the HSMA can still obtain pretty good accuracy, which will significantly help weaken the artificial boundary effect and reduce the size of the simulation domain.
Thus the HSMA can be a useful tool for large-scale simulations of charged systems using the multi-scale hybrid model.

In the future, we plan to combine the hybrid solvent model with solute molecules inside the spherical cavity. One can couple our method with boundary integral equation methods for the simulation of biomolecules~\cite{JBV+:JCP:1991,lu2006order,lu2007new-version-fast-multipole-method,GK:JCP:2013,GGK2019preprint}; or moment/image method for colloidal suspensions~\cite{GJL:JSC:2016,gan2019efficient}. One advantage of the hybrid model lies in the explicit treatment of the electrolyte solvent inside the cavity, thus the ion specific/electrostatic correlation effect can be investigated, and it also avoids the artifacts of PBC~\cite{LXX:NJP:2015}.
Another goal is to implement the HSMA on GPUs to speed up its performance, and we shall also try to apply this multi-scale strategy to particle-based simulations.

\section*{Acknowledgements}
J. F. acknowledge the financial support from the Natural Science Foundation of China (Grant Nos:11571236 and 21773165).
Z. G. was supported by NSF grants DMS-1418966 and DMS-1819094. We want to thank Prof. Zhenli Xu for helpful discussions.


\end{document}